\documentclass[conference,a4paper]{IEEEtran}
\IEEEoverridecommandlockouts

\usepackage{cite}
\usepackage[hyphens]{url}
\usepackage[font=small,labelfont=bf]{caption} 
\usepackage{amsmath,amssymb,amsfonts}
\usepackage{algorithmic}
\usepackage{graphicx}
\usepackage{textcomp}

\usepackage{xcolor}
\usepackage{multirow}
\usepackage{subfigure}
\def\BibTeX{{\rm B\kern-.05em{\sc i\kern-.025em b}\kern-.08em
    T\kern-.1667em\lower.7ex\hbox{E}\kern-.125emX}}

\newcommand{\ie}{\emph{i.e.}, }
\newcommand{\eg}{\emph{e.g.}, }

\usepackage{enumitem}
\usepackage{verbatim}
\usepackage{tikz}

\usetikzlibrary{shapes,arrows,positioning,patterns}

\begin{document}

\title{You Drive Me Crazy! Interactive QoE Assessment for Telepresence Robot Control} 






\author{\IEEEauthorblockN{Hamed Z. Jahromi\IEEEauthorrefmark{1}, Ivan Bartolec\IEEEauthorrefmark{2}, Edwin Gamboa\IEEEauthorrefmark{3}, Andrew Hines\IEEEauthorrefmark{1}, Raimund Schatz\IEEEauthorrefmark{4}}
\IEEEauthorblockA{\IEEEauthorrefmark{1}\textit{School of Computer Science, University College Dublin}, Ireland \\
\IEEEauthorrefmark{2}\textit{Faculty of Electrical Engineering and Computing, University of Zagreb, Croatia} \\
\IEEEauthorrefmark{3}\textit{Institute of Media Technology, TU Ilmenau, Germany } \\
\IEEEauthorrefmark{4}\textit{Center for Technology Experience, AIT Austrian Institute of Technology, Austria}}
hamed.jahromi@ucdconnect.ie,  ivan.bartolec@fer.hr, edwin.gamboa@tu-ilmenau.de,\\ andrew.hines@ucd.ie, raimund.schatz@ait.ac.at}


\IEEEpubid{\makebox[\columnwidth]{978-1-7281-5965-2/20/\$31.00 
\copyright 2020 IEEE \hfill} 
\hspace{\columnsep}\makebox[\columnwidth]{ }}

\maketitle

\begin{abstract}
Telepresence robots (TPRs) are versatile, remotely controlled vehicles that enable physical presence and human-to-human interaction over a distance. Thanks to improving hardware and dropping price points, TPRs enjoy the growing interest in various industries and application domains. Still, a satisfying experience remains key for their acceptance and successful adoption, not only in terms of enabling remote communication with others, but also in terms of managing robot mobility by means of remote navigation.
This paper focuses on the latter aspect of remote operation which has been hitherto neglected. We present the results of an extensive subjective study designed to systematically assess remote navigation Quality of Experience (QoE) in the context of using a TPR live over the Internet. Participants were 'beamed' into a remote office space and asked to perform characteristic TPR remote operation tasks (driving, turning, parking). Visual and control dimensions of their experience were systematically impaired by altering network characteristics (bandwidth, delay and packet loss rate) in a controlled fashion. 
Our results show that users can differentiate well between visual and navigation/control aspects of their experience. Furthermore, QoE impairment sensitivity varies with the actual task at hand. 
\end{abstract}

\begin{tikzpicture}[overlay, remember picture]
\path (current page.north) node (anchor) {};
\end{tikzpicture}
\vspace{-.5cm}
\section{Introduction}
\label{introduction}

Telepresence robots (TPRs) enable their users to socially interact with other people remotely over the internet~\cite{Tsui2015}.
Such interaction not only includes audio and video communication (typically by means of a teleconference system mounted on a mobile robot), but also remote physical presence and mobility ~\cite{Kristoffersson2015}.
TPRs currently enjoy rising popularity and interest due to their potential to be employed in different domains including offices, hospitals, and telemedicine~\cite{Kristoffersson2015,Laniel2017}.
It is their mobility and related remote navigation capabilities that differentiate TPRs from conventional stationary teleconference systems.
Therefore, besides enabling communication, offering a compelling experience during navigation of a TPR is essential for ensuring high user satisfaction.
In this context, the concept of Quality of Experience~(QoE) is relevant since it is ``the degree of delight or annoyance of the user of an application or service", and is affected, among others, by network-related system Influence Factors~(IFs) such as bandwidth, delay, jitter and packet loss~\cite{qualinet2012qoe}.
Subjective studies can be employed to evaluate the impact of different IFs on the QoE of navigating via a TPR.

So far, the experience of tele-navigation has been explored from different points of view.
For instance, in~\cite{Ainasoja2019} three user interface approaches for smartphone teleoperation for TPRs were evaluated.
Moreover, in~\cite{Vlahovic2018} immersion, physical discomfort and overall QoE were studied concerning four locomotion techniques.
Also, Tsui et al.~\cite{Tsui2015} evaluated an augmented reality user interface while a group of users visited a gallery remotely.
Similarly, Kristoffersson et al.~\cite{Kristoffersson2015} evaluated the quality of interaction in a TPR system, assessing factors such as perceived presence and ease of use.
In the context of teleoperating a driving system, variations in latency have been demonstrated to impact users' performance while performing driving tasks in a car simulator system~\cite{Neumeier2019}.
However, to the best of our knowledge, the impact of network-related system IFs on the QoE of navigating TPRs has not been comprehensively investigated yet.
To address this gap, we assess the impact of three network-related system IFs, \ie bandwidth, delay and packet loss rate, on the QoE of navigating a TPR from video and navigation perspectives.
We conducted a subjective QoE study, in which participants performed various navigation tasks with a TPR. The three key contributions of this paper are: 1) quantification of the impact of network IFs on the TPR QoE, 2) empirical assessment to which extent users can discriminate between video and navigation quality while experiencing a multi-sensory experience of TPR navigation,and 3) demonstrate that the TPR QoE sensitivity to network impairments can depend on the actual navigation task performed.  



\section{Background and Related Work}
\label{relatedwork}
A TPR is a wheeled device with wireless Internet connectivity, which allows a human operator to be virtually present and to actively interact in a remote environment via bidirectional audio, video and data transmission~\cite{MacHaret2012, Laniel2017}.
Navigation represents the additional dimension of TPRs when compared with conventional stationary teleconference systems.
Nevertheless, navigating a TPR may pose challenges to the end users that influence their QoE.
Most importantly, user awareness of the navigated environment depends on the quality and variety of the available sensors~\cite{Khan2018}.
For instance, the 2D cameras used in many telepresence systems offer a limited range of vision, which leads to a limited user experience~\cite{Riano2011}.
Moreover, a robot's movement capabilities and responsiveness to interactions impact end-users' experience, too~\cite{Coltin2012}.
Consequently, navigating a robot may result not only in dissatisfaction, but also in undesired accidents such as breaking objects or harming humans.
To study the QoE of navigating a TPR, different IFs should be considered.
QoE IFs are classified into human IFs, system IFs and context IFs~\cite{qualinet2012qoe}.
Regarding the influence of device related system IFs on human IFs, in~\cite{Coltin2012}, three navigation strategies for a TPR were studied. 
The authors found that meeting users' varying preferences may require offering multiple navigation interfaces.
For instance, arrows may be appropriate for small distances and obstacle avoidance, while clicking on the camera image or in a map may be proper for long distance navigation.
Similar results were found in~\cite{Ainasoja2019}. 
Additionally, navigation assistance was studied in~\cite{Riano2011}, in which two TPRs, one including a navigation assistance system, one without, were compared.
They found that such assistance helps users avoid obstacles enhancing their satisfaction.

The user interface design for navigation tasks has also been studied.
\cite{Tsui2015} presents a qualitative study in which the interface ease of use of a TPR was assessed positively by a group of users.
However, the study was conducted in a highly controlled static environment, in which there was only one person and five exhibits of an art gallery used for the study.
Thus, their results are not applicable in dynamic contexts, in which obstacle avoidance may pose a challenge.
Although it was not studied rigorously, the authors found that users required less than half of the time when visiting the environment in person, than during the virtual training.
Regarding context IFs, in~\cite{Kristoffersson2015}, the authors assessed navigation in terms of social interaction and environment layout.
Their results indicate that users found it easy to turn, stop, go backwards, follow a person and go back to the docking station, though high interaction episodes were identified during transitions between different locations.
They found that social interaction may enhance the perceived presence, and the environment layout impacts the perceived ease of navigation.
Navigation challenges have also been studied in other domains such virtual reality. 
Vlahovic et al.~\cite{Vlahovic2018} presented a comparison of four locomotion techniques in a virtual reality environment; \ie controller movement, controller movement with tunneling, teleportation and human joystick.
The overall QoE was rated lower for controller movement and human joystick, which were also the ones which resulted in more physical discomfort for the testers.
The authors found that comfort may have a stronger impact on the QoE for navigation in virtual reality environments than the perceived immersion.

To the best of our knowledge, no study has been conducted so far on how network related System IFs, such as bandwidth, delay and packet loss rate, impact the QoE of navigating a TPR.
Nevertheless, this aspect has been investigated in other domains.
In~\cite{Neumeier2019}, the authors demonstrated that variations in delay impact users' QoE while teleoperating a car in a simulator system. Participants performed navigation tasks in four scenarios: parking, snake, pylon, \ie big double curve and zigzag, and a long track.
Study results indicate that small delays worsen driving performance, and increase perceived workload.
Moreover, they did not find a significant difference between $0.0 s$ and $0.3 s$ delay in the testers' driving performance for the parking scenario.
In the context of TPRs, these results may be different since travel distances are normally shorter and range of vision might be more limited than in car driving scenarios.
To summarise, previous work has demonstrated that 1) navigating a TPR is a challenging endeavour requiring empirical investigation; 2) network related system IFs and type of navigation task may affect QoE of controlling a TPR; and 3) the QoE of navigating a TPR has not been studied considering those system IFs.
Therefore, we conducted a subjective QoE study to investigate the impact of the type of navigation task and network related IFs 
on the QoE of navigating a TPR.

\section{Goals and Methodology}
\label{methodology}

\subsection{Research Goal and Questions}
\label{sec:method:rq}

The goal of our study is to examine the impact of bandwidth, delay and packet loss rate impairments on the operator's QoE when using a TPR in an office context, in which users utilise the TPR to navigate to different parts of a remote office. 
We chose the office context since it is one of the most common and thus representative scenarios for wheeled TPR usage~\cite{beno2018TPRsRemote}.
To conduct the study, we formulate the following research questions: \textbf{RQ1})  How do network IFs (bandwidth, delay and packet loss rate) impact QoE for TPR navigation? \textbf{RQ2}) To what extent do users discriminate between navigation task quality and video streaming quality? and \textbf{RQ3}) Does type of task exert significant influence on TPR QoE?
Note that, the RQ2 and RQ3 are not specific to the types of influence factors.
\subsection{Study Process}

Study participants were situated in a laboratory setting in Dublin, Ireland, where they navigated a TPR via keyboard input on a desktop computer. The TPR was placed in a real office setting of California Telecom company in California, USA. 
The TPR was connected to the office Wi-Fi connection, which was unimpaired since the network parameters were manipulated only on the user side.
Prior to the experiment, the participants were screened for correct visual acuity using Snellen charts~\cite{snellen} and for color vision using Ishihara charts\cite{ishihara}. All participants had a short training at the beginning of the test session. They familiarised with the course of the study, the measured parameters, the TPR system, and the controls used for different tasks. Then, the participants were asked to fill a pre-test questionnaire which included general questions such as age, gender, and prior TPR, Virtual Reality (VR) and gaming experience. The study consisted of 23 test scenarios, each featuring a specific network condition and a specific navigation task. Network conditions and navigation tasks will be discussed in detail in the subsequent subsections. Each participant was exposed to all 23 test scenarios in a randomised order. More specifically, network impairments were randomised while navigation tasks followed a certain sequence (\ie firstly, a TPR goes from location A to location B, secondly, the TPR has to rotate, and thirdly, it goes from B to A) for logistical reasons. After finishing each test scenario, \ie completing a task under defined network connection parameters, participants were asked to rate their perceived quality of navigation ($q1$), and quality of video ($q2$) using the Absolute Category Rating (ACR) method. Also, they indicated binary acceptability in terms of whether they would use the TPR in real-life everyday situations given the experienced conditions ($q3$). To this end, we used a Web-based digital questionnaire based on TheFragebogen\footnote{\url{http://www.thefragebogen.de} [last accessed: Jan 27, 2020]}.

\subsection{Telepresence Robot Design}
\begin{figure}[ht]
  \centering
  \subfigure[Design of Beam TPR (taken from \cite{beampro}).]{\includegraphics[scale=0.23]{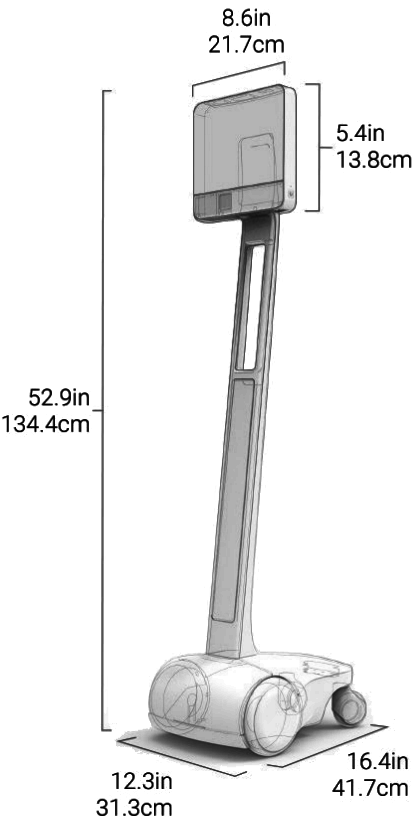}
  \label{fig:beampro}}\quad
  \subfigure[Beam client application user interface.]{\includegraphics[scale=0.25]{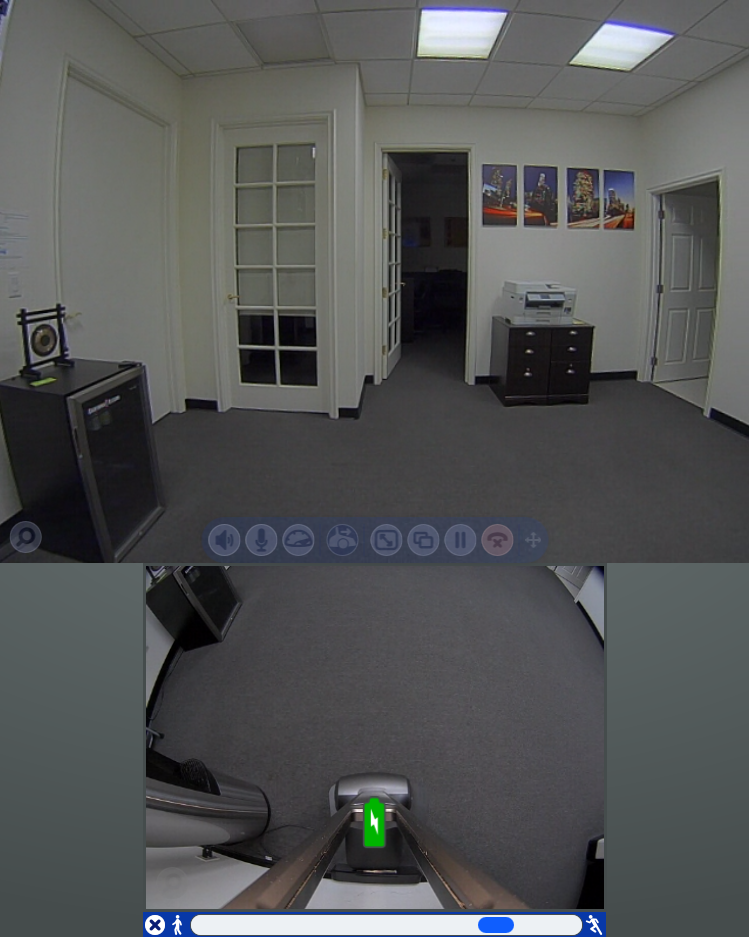}
  \label{fig:ui}}
  \caption{Hardware and software of a Beam TPR.}
  \label{fig:beampro_ui}
  \vspace*{-4mm}
\end{figure}

Figure~\ref{fig:beampro} shows the design of the ``Beam'' TPR used in the study. It is composed of a fixed head that contains an LCD display and two cameras (a front camera and a wide-angle view camera), a body made out of aluminium and steel, and a differential five-wheeled chassis. The chassis offers stability and mobility constrained to the following motions: driving forward, driving backwards, and rotation. The TPR has an option of assisted navigation where speed is adjusted according to surroundings to reduce collisions. However, this option was not enabled during our experiments to avoid unintended influences on participants' experience.

\subsection{User Interface}
The TPR enables the user, \ie robot operator, to be virtually present from a remote location via a client application. The user interface (UI) contains the camera feed from the two aforementioned cameras, the user device camera feed, and a set of settings for controlling TPR performance and video quality. The user sends commands to the TPR by pressing buttons or keys from a keyboard, mouse, or joystick. For the purpose of the experiment, users are presented with the default UI design as shown in Fig.~\ref{fig:ui}, although a full screen display of the camera feed may also be used for better visibility in real-life scenarios. Default values are used for all settings.


\subsection{Task Design}

\begin{figure}[ht]
  \centering
  \subfigure[Depiction of the office environment and tasks employed for the study.]{\includegraphics[scale=0.20]{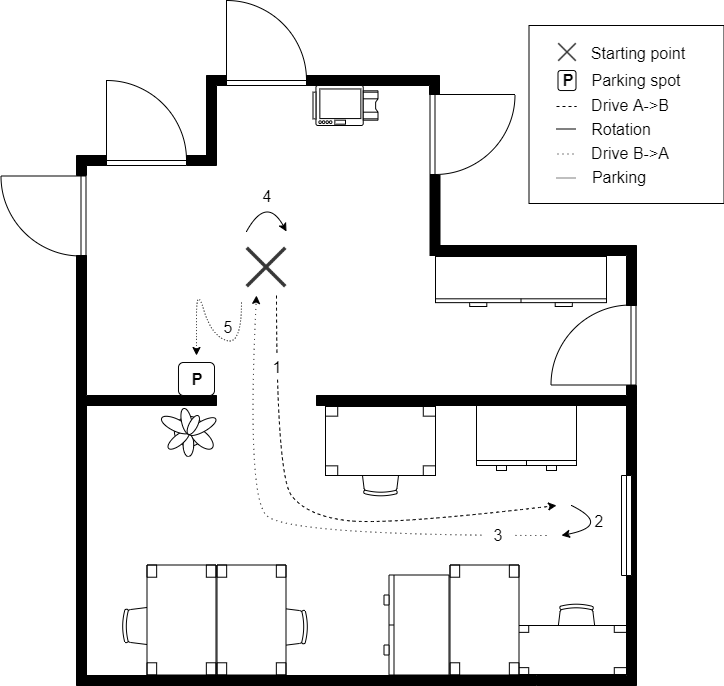}
  \label{fig:blueprint}}\quad
  \subfigure[Rotation task depicted.]{\includegraphics[scale=0.22]{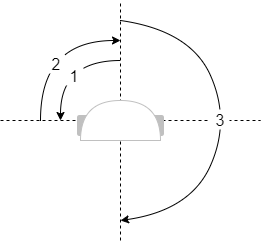}
  \label{fig:rotation}}
  \caption{Office environment and tasks depiction.}
  \label{fig:blueprint_rotation}
  \vspace*{-5mm}
\end{figure}

Under controlled (impaired or unimpaired) network conditions,  participants performed tasks belonging to three categories. In the first task category, they navigated the TPR from a start position to an end position in a different room ($drive$ task). While navigating from the start to the end position, they avoided obstacles commonly found in an office environment such as desks, chairs, bins, etc. In the second task category $turn$, the participants performed a rotation motion around the vertical axis (\eg only by using turn left and turn right keys) as depicted in Fig.~\ref{fig:rotation}. That is, the TPR had to rotate 90 degrees to the left. Then, it had to rotate 90 degrees to the right. Next, it had to rotate 180 degrees to the right and face the direction from which it arrived. After finishing the rotation task, participants navigated back from end position to start position ($drive$), and once arrived, they repeated a rotation task. Finally, in the third category $park$, the participants parked the TPR on its charging station starting from the start position. This was the only task which encouraged participants to drive backwards. The different tasks are illustrated in Fig.~\ref{fig:blueprint}.


\subsection{Network Conditions}
Bandwidth, delay and packet loss rate were manipulated on the local PC by using \textit{dnctl} and \textit{pfctl}. Each network IF has four associated values to emulate different levels of quality as presented in Table~\ref{tab:netconditions}. The specific network emulator settings were selected after conducting multiple pre-study sessions. Due to the geographical distance between participants and the TPR, the connection already had a base delay value (\ie approximately 200 \textit{ms}). 
The participants $drive$ and $turn$ a TPR under each impaired network IF from level 1, 2, 3 and 4. The best quality condition, presented here as level 4, did not feature added delay or packet loss rate, only a bandwidth cap set to 4 \textit{Mbps} which exceeds the 3 \textit{Mbps} required by the system as optimal technical conditions. This level 4 was used only once per task category as a hidden reference. Parking task was only performed with added delay conditions, since bandwidth, and packet loss rate only caused lower noticeable effect during the pre-study session for this activity. More specifically, parking tasks were performed under impaired delay values from levels 1, 2, and 4. This resulted in 23 scenarios in total for each participant with each non-reference scenario featuring one impaired network parameter.
Note that, since the TPR was operated remotely over the Internet, we ensured stable experimental conditions by using dedicated access lines on both sites (local laboratory and remote office) and constantly monitoring the three network QoS parameters during the experiments.

\begin{table}[ht]
\begin{center}
\caption{Target values for the factor levels used to implement the different network conditions used in the test scenarios.}
\label{tab:netconditions}
\begin{tabular}{llll}
\hline
\multirow{2}{*}{Level} & \multicolumn{3}{c}{Values}                         \\
\cline{2-4}
                       & Bandwidth (\textit{Mbps}) & Delay (\textit{ms}) & Packet loss rate (\%) \\
\hline
Low (1)                     & 0.50             & 700          & 5            \\
Medium (2)                  & 0.90             & 450          & 2            \\
High (3)                    & 1.25             & 350          & 1            \\
Best (4)                    & 4.00             & 200      & Not set      \\
\hline
\end{tabular}
\renewcommand{\arraystretch}{1}
\end{center}
\vspace*{-4mm}
\end{table}






\section{Experimental Results}
\label{results}
\begin{figure}[!tp]
\centering
\includegraphics[clip,width=0.49\columnwidth]{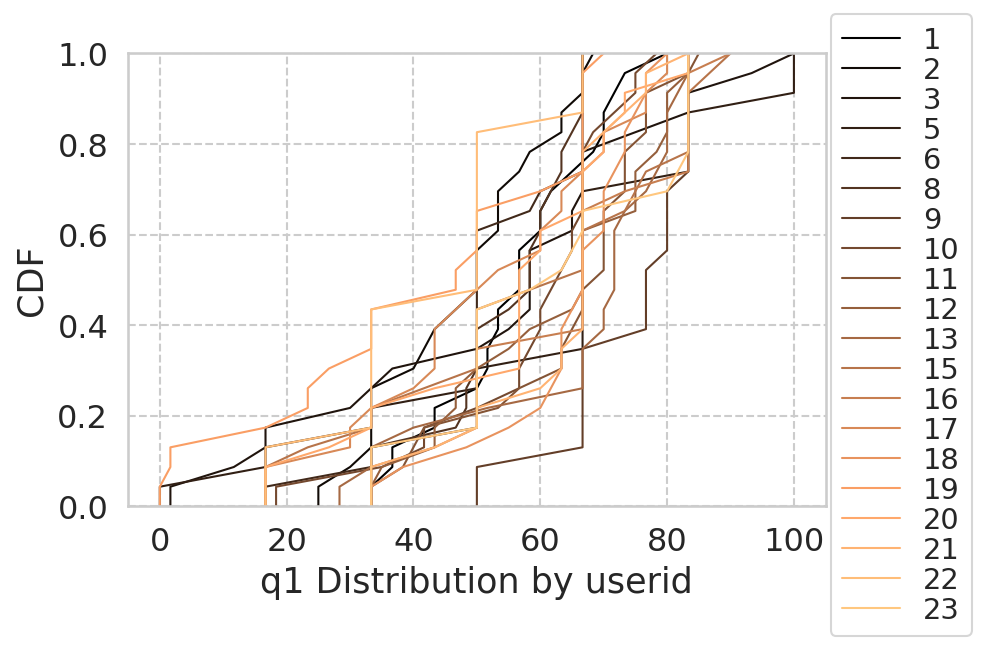}
\includegraphics[clip,width=0.49\columnwidth]{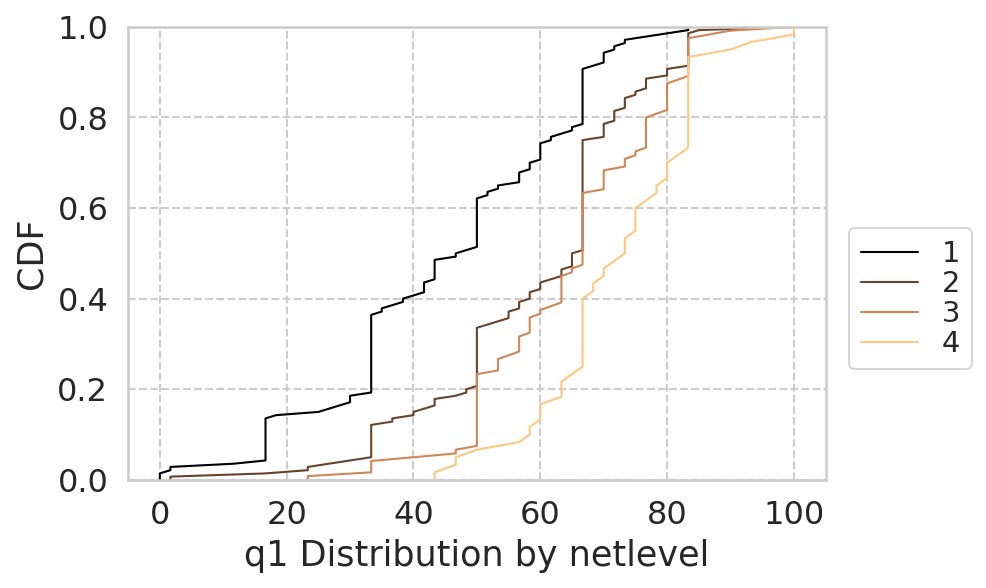}
\includegraphics[clip,width=0.49\columnwidth]{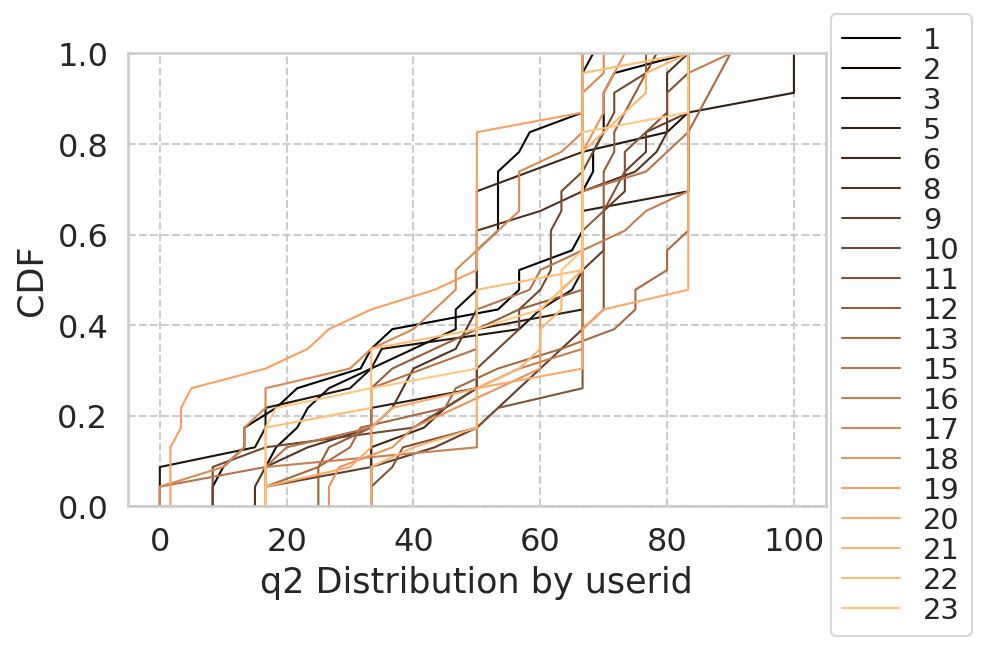}
\includegraphics[clip,width=0.49\columnwidth]{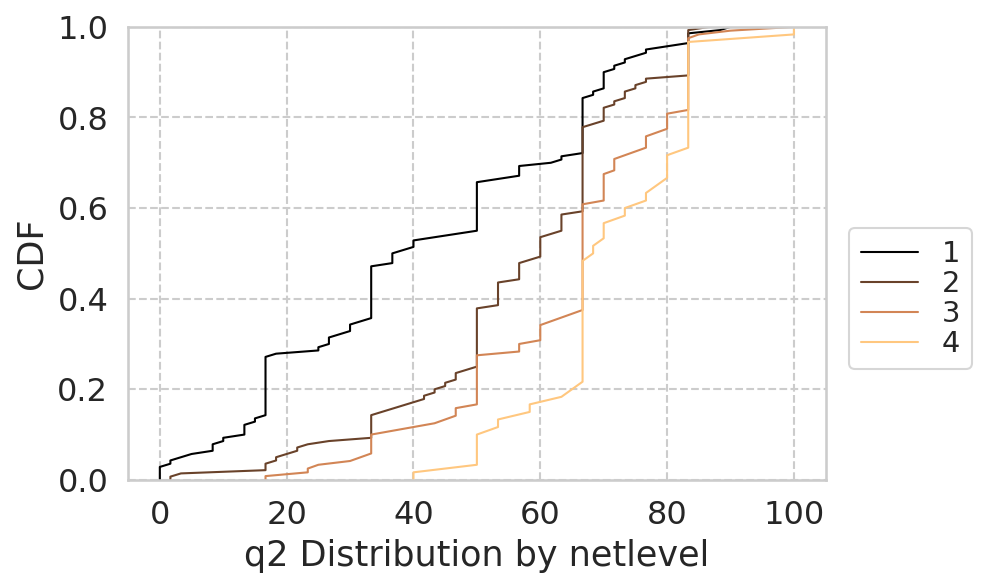}
\caption{The cumulative distribution function (CDF) plots of TPR navigation QoE scores. $q1$ ratings grouped by user (top left), $q1$  network quality level (top right), $q2$ ratings grouped by user (bottom left), $q2$  network quality level (bottom right).}
\label{fig:ss_cdf}
\vspace*{-5mm}
\end{figure}

A total of 23 subjects participated in our QoE lab study: 8 subjects were female and 15 male, with an average and median age of 30 years old. All subjects had normal or corrected eye vision. The survey of subjects explored prior experience with immersive and interactive technologies: TPRs; gaming and VR. 69\% of participants were unfamiliar with TPRs; 22\% had heard of them but not used them and 4\% had a small amount of prior experience (tried once or twice).
Regarding gaming experience, 30\% of subjects reported playing computer games regularly, i.e. more than three times a month, 34\% less than three times a month and 26\% once or twice. The remaining 8\% did not have any previous gaming experience. 69\% of participants had tried VR technology once or twice with 21\% and 8\% just heard about VR technologies or never heard of it, respectively. The post-test surveys did not report task boredom as a factor, In fact, people enjoyed the experience.

Surprisingly, our statistical analysis found no significant impact of previous experience, gender or age of subjects on their ratings. We can only hypothesise that higher participant numbers could yield more conclusive results on this aspect. 

As part of our data preprocessing following ITU-T BT.500, we detected three participants as outliers and excluded them from our subsequent analysis. 
For the remaining dataset ($N=20$), we generated CDF (cumulative distribution function) plots to depict the distribution of the user's ratings and the rating behaviour (see Fig.~\ref{fig:ss_cdf}). The left subplots (distribution by $userid$) reveal a fair level of diversity between participants in terms of scoring. Score distributions by $netlevels$ (right subplots) demonstrate that the different network conditions (levels: 1-4) have visible impact on TPR navigation and video QoE ratings, confirming that our measurement setup is sensitive. 

Figure~\ref{fig:mos_all} shows the influence of different network conditions; \ie bandwidth ($bw$), packet loss rate ($plr$), and $delay$ on the perceived quality as rated by the participants Mean Opinion Score (MOS) for a given task. The top bar plot displays the MOS of users' navigation quality ratings ($q1$) while the middle plot depicts the MOS of users' video quality ratings ($q2$). 
The third subplot (bottom) depicts the ratios of subjects' answer to the binary service acceptability question ($q3$), seemingly correlated with $q1$ and $q2$. The remainder of this section answers the three research question (see section \ref{sec:method:rq}), including underlying analysis.


\subsection{Network Impairments and TPR Navigation QoE (RQ1)}
All three network IFs have significant impact on navigation QoE, but in different ways. $bw$ shows no influence on navigation QoE until a threshold of 0.9 Mbps while 1\% $plr$ has noticeable impact on the navigation QoE.
Statistical analysis (Table~\ref{tab:mm_anova} and Fig.~\ref{fig:subj_lsmeans}) shows that the impact of the different network impairments ($bw$, $delay$ and $plr$) on navigation quality ratings ($q1$) is significant ($pval.q1 <0.001$). This is in line with Fig.~\ref{fig:mos_all} (top, $q1$) showing that the absolute levels of navigation QoE MOS (as well as their sensitivity to network impairment level) depend on the actual impairment type. 


\subsection{Discrimination between TPR Video Quality and Navigation Quality (RQ2)}
\begin{figure}[!tp]
\centering
\includegraphics[clip,width=\columnwidth]{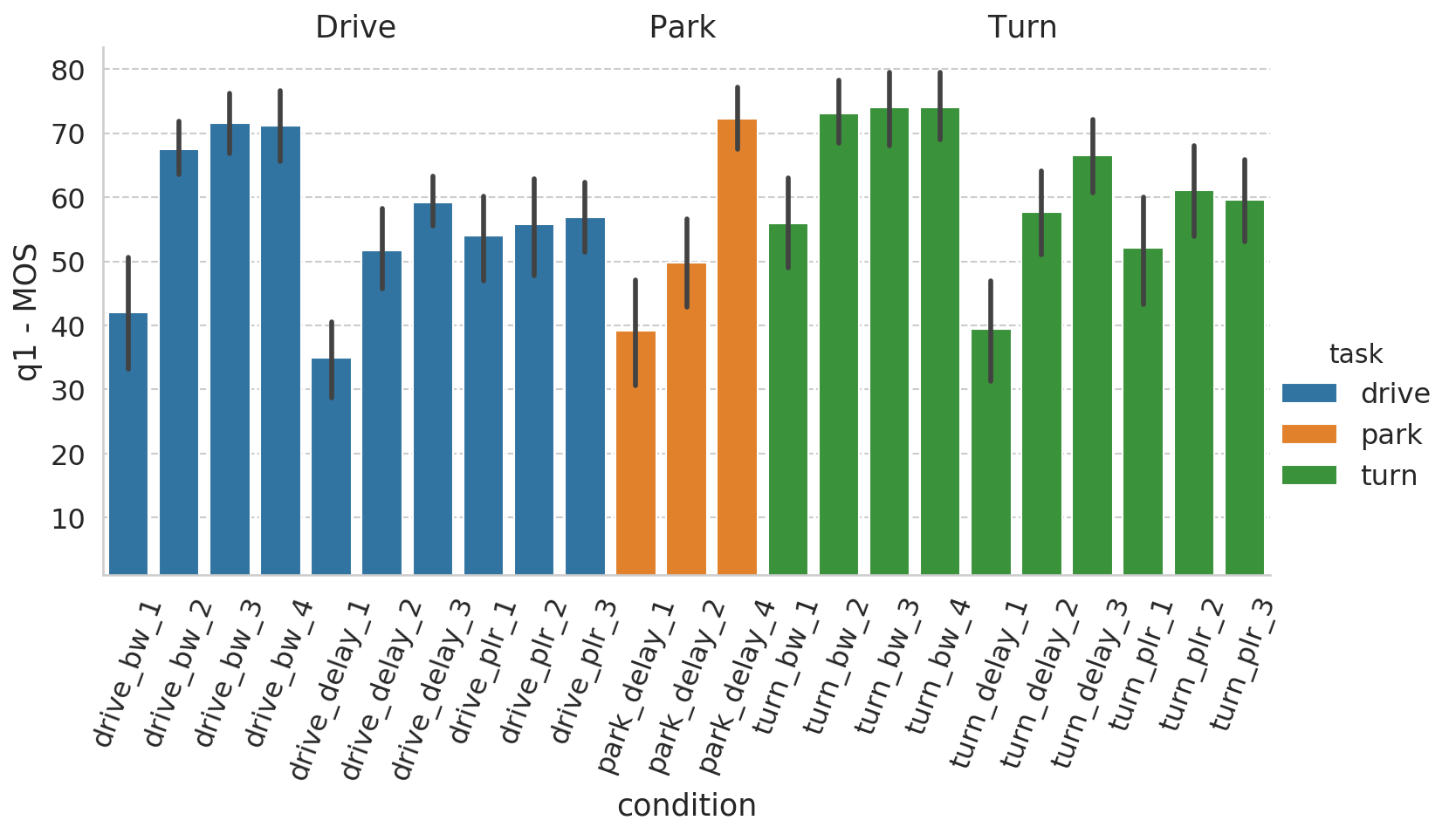}
\includegraphics[clip,width=\columnwidth]{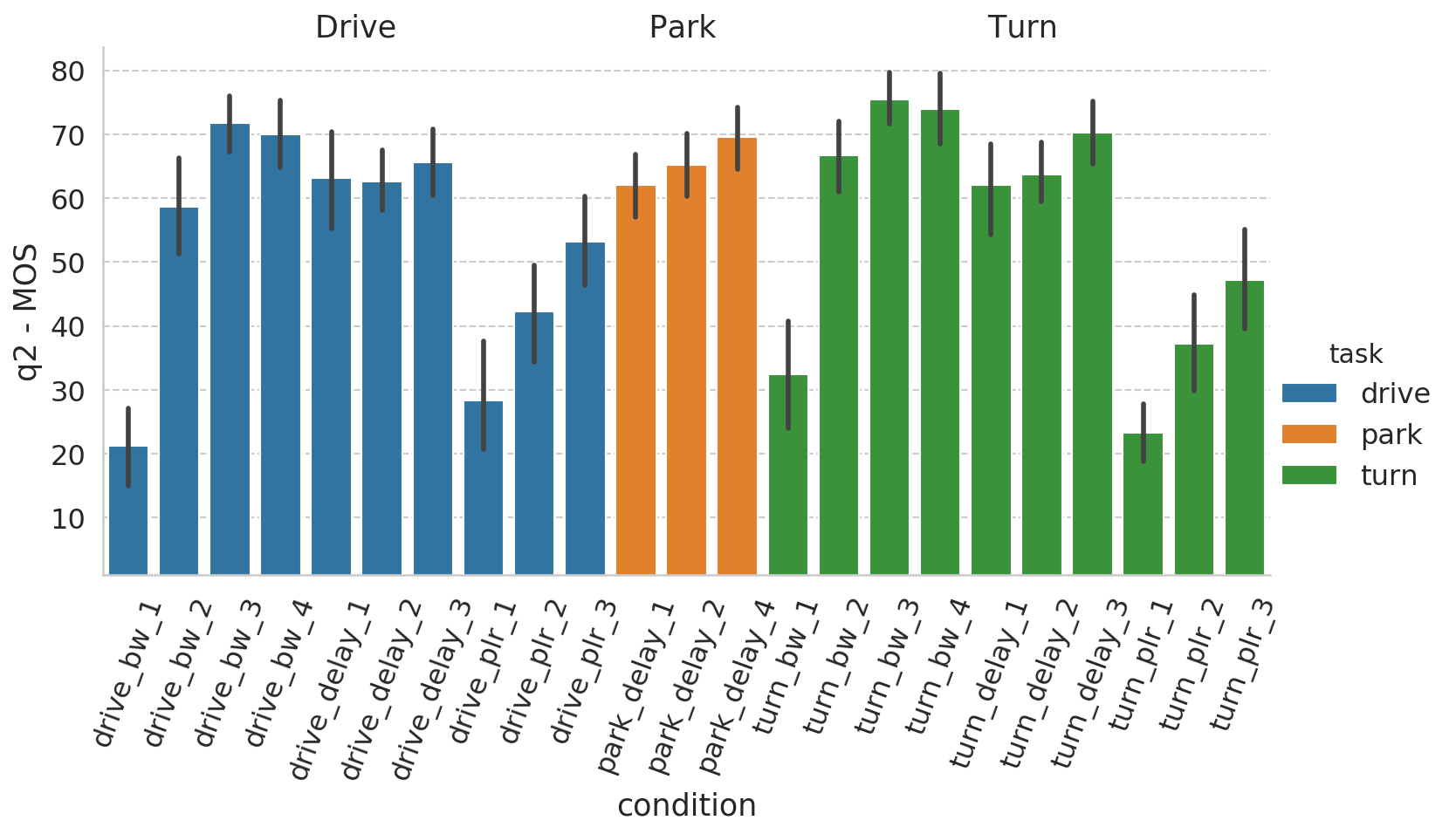}
\includegraphics[clip,width=\columnwidth]{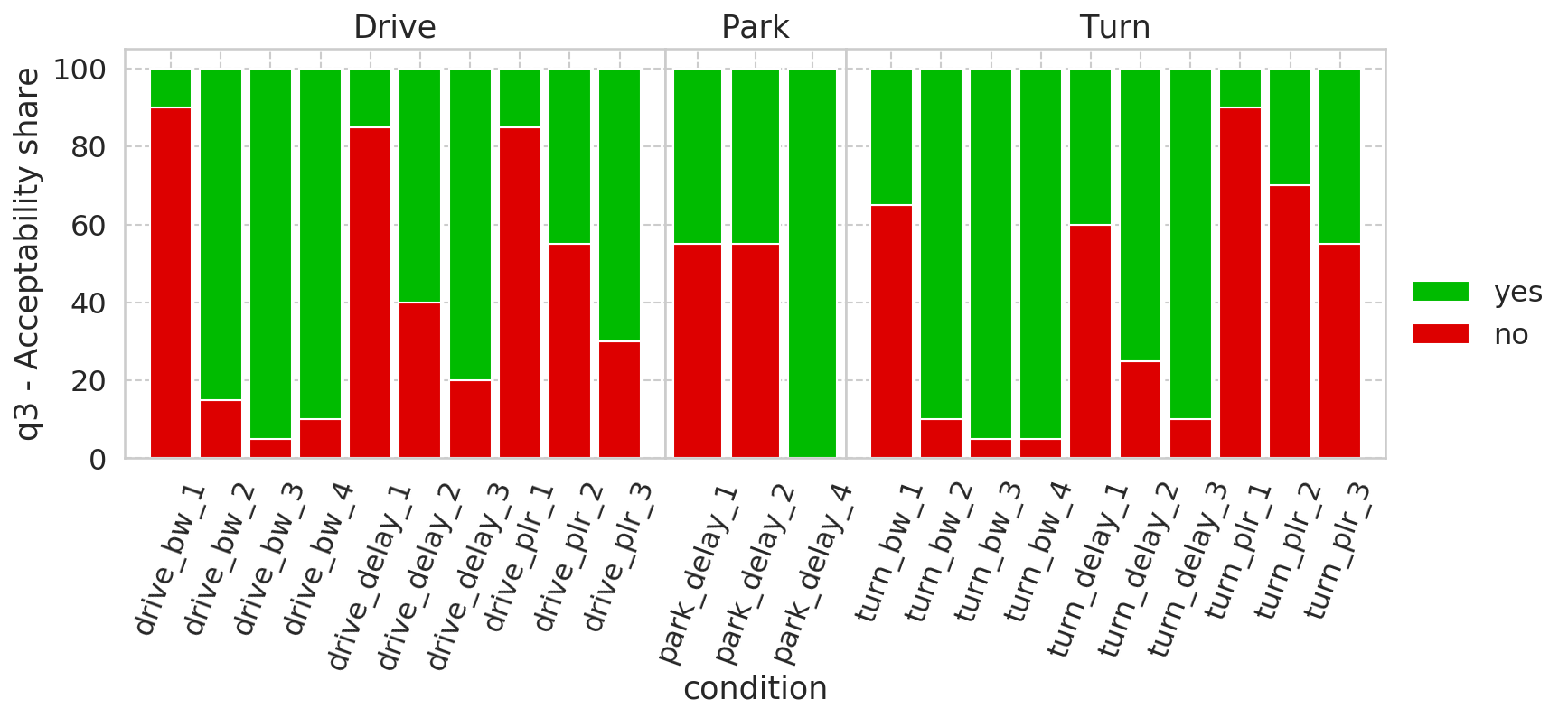}
\caption{MOS for navigation ($q1$, top) and video quality ($q2$, middle) for all test conditions of the study. MOS is normalised to 1-100, with 1-20 equating ``bad" and 81-100 equating ``excellent" quality. (MOS CI = 95\%). Each color group designates a task, with conditions being further grouped by network emulation parameter, ordered by increasing technical quality level of that parameter (see Table~\ref{tab:netconditions}).
Acceptability results ($q3$, bottom) are displayed as yes vs. no proportions. Conditions with level 4 are (hidden) reference conditions.
Condition code format: task\_parameter\_level.}
\label{fig:mos_all}
\vspace*{-5mm}
\end{figure}

The subjects are capable of discriminating between video quality and navigation quality, \ie they treat them as separate concepts when it comes to experience assessment.  
Our Mixed Model ANOVA analysis (Table~\ref{tab:mm_anova}) shows that the impact of $bw$ and $plr$ on TPR video quality ratings ($q2$) is significant ($pval.q2<0.001$). However, $delay$ does not exert significant influence on TPR video QoE. This is not surprising, since packet delay, when not at excessive levels, generally does not affect video quality. 
Similar to navigation QoE, the actual degree of influence of network impairments on perceived video quality varies with impairment type. However, a comparison of $q1$ and $q2$ subplots (Fig.~\ref{fig:mos_all}) shows that changes in MOS across different impairment levels diverge between the two in terms of amplitude. 
To quantify this divergence as the extent to which $q1$ and $q2$ ratings are not aligned, we performed Spearman Rank Ordered Correlation Coefficient (SROCC) analysis, revealing a weak correlation between $q1$ and $q2$ ($SROCC = 0.47$).  The service acceptability $q3$ actually correlates strongly with the geometric mean of $q1$ and $q2$ ($SROCC=0.78$) and less with each individual QoE dimension ($SROCC=0.64$ and $0.69$ for $q1$ and $q2$ respectively). This suggests that acceptability equally depends on delivering on both dimensions, video and navigation QoE, which were affected to varying extent throughout the experiment.
Together with the above ANOVA and SROCC results, this leads us to conclude that in the context of TPR QoE, subjects can discriminate between video quality and navigation quality.

\subsection{Influence of Task on QoE (RQ3)}
\begin{figure}[!tp]
\centering
\includegraphics[clip,width=\columnwidth]{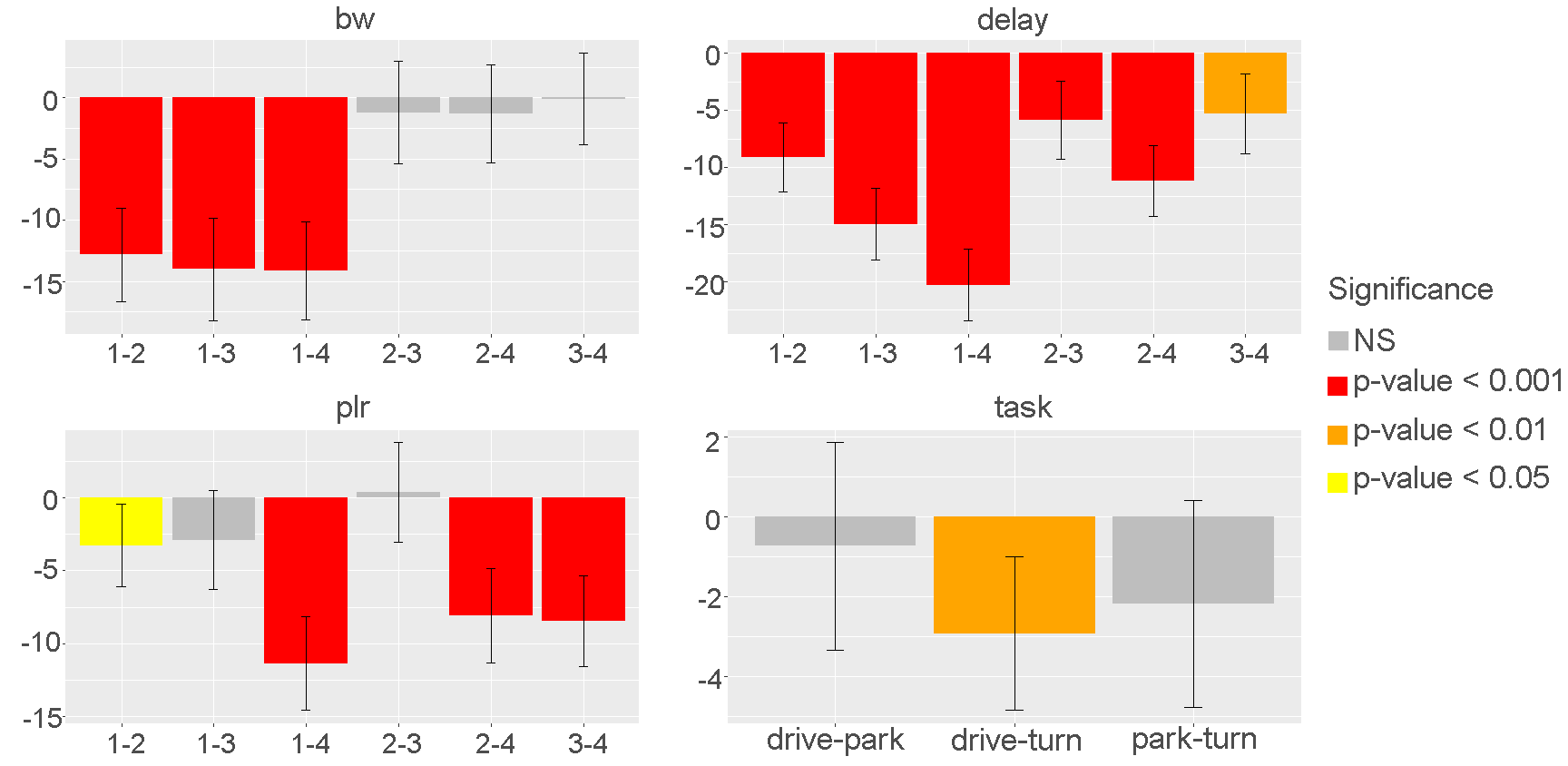}
\caption{Post-hoc Analysis: Differences of Least-Square Means Analysis for $q1$ for significant study factors. Coloured bars indicate (bonferroni-corrected) significant differences between factor levels.}
\label{fig:subj_lsmeans}
\vspace*{-4mm}
\end{figure}

The type of TPR task has more impact on the navigation QoE than the streaming video QoE.  
The results of our statistical analysis (see Table~\ref{tab:mm_anova} and Fig.~\ref{fig:subj_lsmeans} bottom right) suggest that the actual task affects QoE sensitivity, depending on network impairment type. For example, the interaction between bandwidth and task is significant for navigation QoE ($bw:task$, $pval.q1<0.01$, see Table~\ref{tab:mm_anova}, fixed effects), which means that changes in bandwidth were rated differently depending on task type. On the other hand, this was not the case for delay ($delay:task$) and packet loss rate ($plr:task$). 

Regarding video quality ($q2$), we do not see a significant impact of task on QoE sensitivity, except for the borderline case for packet loss rate ($plr:task$, $pval.q2=0.05$). 
Thus, we can assume that the navigation QoE is more sensitive to task type than the video QoE. We explain this by our observation, that navigation per se is more context- (and thus task-) dependent than pure video quality assessment. 
Note, that repeated execution of the same task (represented by factor $iter$) does not exert significant influence on $q1$ or $q2$ (see Table~\ref{tab:mm_anova}, fixed effects). This suggests that in our study, boredom or learning effects did not significantly affect QoE ratings.

\begin{table}[ht]
\centering
\begin{tabular}{rrrr@{\hskip 0.1cm}lr@{\hskip 0.1cm}l}
  \hline
Fixed & Fval.q1 & Fval.q2 & pval.q1 & & pval.q2 &  \\ 
  \hline
bw & 17.442 & 68.087 & $<$0.001 & *** & $<$0.001 & *** \\ 
  delay & 14.184 & 0.681 & $<$0.001 & *** & 0.565 &   \\ 
  plr & 7.211 & 38.433 & $<$0.001 & *** & $<$0.001 & *** \\ 
  task & 2.550 & 0.551 & 0.08 &   & 0.577 &   \\ 
  iter & 0.502 & 2.126 & 0.734 &   & 0.079 &   \\ 
  bw:delay & n.a. & n.a. & 1.000 &   & 1.000 &   \\ 
  bw:plr & n.a. & n.a. & 1.000 &   & 1.000 &   \\ 
  bw:task & 3.904 & 1.575 & 0.01 & * & 0.196 &   \\ 
  bw:iter & 0.102 & 0.736 & 0.991 &   & 0.599 &   \\ 
  delay:plr & n.a. & n.a. & 1.000 &   & 1.000 &   \\ 
  delay:task & 1.985 & 1.031 & 0.099 &   & 0.392 &   \\ 
  delay:iter & 1.620 & 1.420 & 0.172 &   & 0.229 &   \\ 
  plr:task & 1.158 & 2.649 & 0.328 &   & 0.05 &   \\ 
  plr:iter & 1.542 & 1.817 & 0.208 &   & 0.146 &   \\ 
  task:iter & 0.561 & 0.162 & 0.761 &   & 0.986 &   \\  
    \hline
Random & Chi.q1 & Chi.q2 & pval.q1 & & pval.q2 &  \\ 
  \hline
bw:userid & 14.813 & 3.039 & $<$0.001 & *** & 0.081 &   \\ 
  delay:userid & 0.000 & 1.554 & 1.000 &   & 0.213 &   \\ 
  plr:userid & 0.000 & 2.255 & 0.999 &   & 0.133 &   \\ 
  task:userid & 4.417 & 0.000 & 0.036 & * & 0.995 &   \\ 
  iter:userid & 0.000 & 0.000 & 1.000 &   & 1.000 &   \\ 
  bw:delay:userid & 0.052 & 0.000 & 0.82 &   & 0.984 &   \\ 
  bw:plr:userid & 0.000 & 0.006 & 1.000 &   & 0.936 &   \\ 
  bw:task:userid & 0.000 & 0.000 & 0.985 &   & 0.998 &   \\ 
  bw:iter:userid & 0.783 & 5.580 & 0.376 &   & 0.018 & * \\ 
  delay:plr:userid & 1.208 & 0.000 & 0.272 &   & 0.997 &   \\ 
  delay:task:userid & 0.000 & 0.002 & 0.998 &   & 0.966 &   \\ 
  delay:iter:userid & 10.803 & 14.658 & 0.001 &   & $<$0.001 & *** \\ 
  plr:task:userid & 0.017 & 1.537 & 0.895 &   & 0.215 &   \\ 
  plr:iter:userid & 6.307 & 0.000 & 0.012 & * & 1.000 &   \\ 
  task:iter:userid & 1.023 & 0.000 & 0.312 &   & 1.000 &   \\ 
  userid & 2.062 & 4.196 & 0.151 &   & 0.041 & * \\ 
   \hline
\multicolumn{5}{l}{\scriptsize{{*** }$p<0.001$, {** }$p<0.01$, {* }$p<0.05$}}
\end{tabular}
\caption{Mixed-Model ANOVA Results for Fixed (F-Test) and Random Effects (Likelihood-Ratio Test). Target variables: q1, q2. Asterisks indicate levels of statistical significance.} 
\label{tab:mm_anova}
\vspace*{-5mm}
\end{table}

\section{Discussion}
\label{sec:discussion}
Our subjective assessment of TPR QoE has revealed various facets (navigation, video, task sensitivity, etc.) of TPR QoE. We thus consider the results of our study to be relevant for TPR manufacturers, network QoE management orchestrators and professional users. The TPR user has a multi-sensory QoE experience for navigation with keyboard input and visual feedback via the video stream. Our results show that $delay$ influences navigation QoE ($q1$) but not video QoE ($q2$). As the video stream and navigation command communications are not in direct competition for network resources -- changing the video resolution or frame rate may not influence navigation command responses. This suggests that changing the video resolution/frame rate may be of limited benefit, but that sensory alignment cues, e.g. video cues to visually confirm that commands have been received, may have more of an effect on navigation QoE. 
We believe that the experimental approach and setup presented here not only is a good starting point for other TPR QoE studies, but also might be transferred to QoE assessment in similar application domains of cyber-physical systems \cite{Hammer_2018}. Our set of navigation tasks (drive, turn, park) proved to be useful, since they not only provided realistic scenarios and reduced user boredom, but also succeeded in triggering different levels of QoE sensitivity. Also, the training phase was an essential component of the experimental process. It allowed participants to  familiarise themselves with the TPR control and to perform the tasks prior to undertaking the experiment, countering potential usability issues and novelty effects. Still, the ANOVA analysis (Table~\ref{tab:mm_anova}) shows the presence of a few significant random effects that indicate diversity among participants in terms of perception and rating that we want to explore by further analysis and experiments.

\section{Conclusion and Future Work}
\label{conclusion}

In this paper, we evaluated the influence of network impairments (bandwidth, delay and packet loss rate) on TPR QoE on behalf of a subjective experiment addressing visual quality, navigation and acceptability aspects. Our experiment features a very common TPR use case (remote presence in an enterprise office) and sheds light on the QoE impact of network-related impairments, task and other factors in this context.
Our results show that users can differentiate fairly well between visual and navigation/control aspects of TPR operation. 
In general, visual and navigation QoE sensitivity to specific impairments correlate weakly with each other, depending on the actual task at hand. 
Our study was intentionally limited in scope by focusing on QoE assessment of remote TPR navigation, omitting other aspects such as social interaction, presence and embodiment.
Also, the three considered IFs are independently considered in our methodology. However, in practice, they may jointly affect QoE. Thus, concerning future work we foresee the development of a more generic, empirically validated TPR QoE framework, taking additional aspects such as human-to-human and human-to-robot interaction into account. To this end, we also plan to investigate the viability of using QoE models and frameworks from other related domains like VR and gaming  to better characterise and predict TPR QoE. 




\section{Acknowledgement}
The authors would like to thank California Telecom Inc. and ECML/PKDD Summer School 19 for supporting this research. This publication has also emanated from research supported in part by a research grant from Science Foundation Ireland and is co-funded under the European Regional Development Fund under Grant Number 13/RC/2077 and Grant Number SFI/12/RC/2289\_P2. I. Bartolec's work has been supported in part by the Croatian Science Foundation under the project IP-2019-04-9793 (Q-MERSIVE).

\Urlmuskip=0mu plus 1mu\relax
\bibliographystyle{./bibliography/IEEEtran}
\bibliography{main.bib}

\vspace{12pt}
\end{document}